\documentclass[11pt]{article}

\pdfoutput=1

\usepackage[T1]{fontenc}
\usepackage[latin9]{inputenc}
\usepackage[a4paper]{geometry}
\usepackage[active]{srcltx}
\usepackage{amsmath}
\usepackage{amssymb}
\usepackage{esint}
\usepackage{ulem}

\makeatletter
\usepackage{textcomp}

\pdfoutput=1 % if your are submitting a pdflatex (i.e. if you have
\usepackage{jheppub}% for details on the use of the package, please
\usepackage{etoolbox}% http://ctan.org/pkg/etoolbox
    
    \patchcmd{\maketitle}{\@fpheader}{}{}{}

\newcommand*\xbar[1]{%
  \hbox{%
    \vbox{%
      \hrule height 0.5pt % The actual bar
      \kern0.3ex%         % Distance between bar and symbol
      \hbox{%
        \kern-0.0em%      % Shortening on the left side
        \ensuremath{#1}%
        \kern-0.0em%      % Shortening on the right side
      }%
    }%
  }%
}

\usepackage{amsfonts}

\usepackage{bbm}

\newcommand{\be}{\begin{equation}}
\newcommand{\ee}{\end{equation}}
\newcommand{\bea}{\begin{eqnarray}}
\newcommand{\eea}{\end{eqnarray}}

\title{\boldmath The final Kasner regime inside black holes with scalar or vector hair}
\author[a,b]{Marc Henneaux}

\affiliation[a]{Universit\'e Libre de Bruxelles and International Solvay Institutes, ULB-Campus Plaine CP231, B-1050 Brussels, Belgium}
\affiliation[b]{Coll\`ege de France, 11 place Marcelin Berthelot, 75005 Paris, France}

\emailAdd{marc.henneaux@ulb.be}

\preprint{}

\abstract{The final (close to the singularity) dynamical behavior of the metric inside  black holes with massive charged scalar or vector hair is analyzed for general anisotropic and inhomogeneous initial conditions.  These solutions are relevant to a holographic realization of superconductivity.   It is shown that the dynamics falls within the scope of the ``cosmological billiard'' description and that in both cases, the corresponding hyperbolic billiard region has infinite volume so that the system ultimately settles down to a final Kasner regime.  For massive vector hair, the conclusion holds because the longitudinal mode  plays the same role as a scalar field. There exists, however, a measure-zero subset of solutions characterized by vanishing longitudinal modes that exhibit a chaotic behavior with an infinite number of BKL oscillations as one goes to the singularity.}

\makeatother

\begin{document}
\maketitle \flushbottom

 \newpage{}

\section{Introduction}

The study of the internal structure of charged black holes with scalar or vector hair has attracted recently a considerable attention \cite{Hartnoll:2020rwq,Hartnoll:2020fhc,Cai:2020wrp,VandeMoortel:2021gsp,Dias:2021afz,Cai:2021obq}.  These solutions  are relevant to the holographic realization of superconductors \cite{Gubser:2008px,Hartnoll:2008vx,Hartnoll:2008kx,Gubser:2008wv,Cai:2013aca} and are described by the Einstein action with negative cosmological constant coupled to the sum of the Maxwell action  and the action for massive charged scalar or vector fields. Specifically, the following cases have been considered,
\be
S = \int d^D x \sqrt{-g} ( R - 2 \Lambda - \frac14 F_{\mu \nu} F^{\mu \nu} + L^c) \label{eq:LTotal}
\ee
with 
\be
L^c = - g^{\mu \nu} D_\mu \phi^\dagger D_\nu \phi - m^2 \phi^\dagger \phi - \alpha \phi^\dagger \phi  F_{\mu \nu} F^{\mu \nu} \label{eq:LScalar}
\ee
(charged massive scalar field $\phi$ with complex conjugate $\phi^\dagger$) or
\be
L^c = - \frac12 \rho_{\mu \nu}^\dagger \rho^{\mu \nu} - m^2 g^{\mu \nu} \rho_\mu \rho_\nu   \label{eq:LVector}
\ee
(charged massive vector field $\rho_\mu$ with complex conjugate $\rho_\mu^\dagger$).  Here, $\Lambda$ is the cosmological constant,  the covariant derivatives are
\be
D_\mu \phi = \partial_\mu  \phi - iq A_\mu \phi, \qquad D_\mu \rho_\nu = \nabla_\mu \rho_\nu - iq A_\mu \rho_\nu, \qquad (\nabla _\mu \rho_\nu = \partial_\mu \rho_\nu - \rho_\lambda\Gamma^\lambda_{\; \; \nu \mu} ),  
\ee
 while $\alpha$ is a coupling constant for the non-minimal coupling term $\phi^\dagger \phi  F_{\mu \nu} F^{\mu \nu}$\footnote{The  non-minimal coupling term  $iq \gamma \rho_\mu \rho_\nu^\dagger F^{\mu \nu}$ has  also been included in the vector case (\ref{eq:LVector}) by the authors of  \cite{Cai:2013aca}.   We shall briefly comment on it in the conclusions. }.

In the absence of the charged matter fields ($\phi$, $\rho_\mu$), the solution is the Reissner-N\"ordstrom AdS black hole, with an inner Cauchy horizon and a timelike singularity. However, as shown  in \cite{Hartnoll:2020rwq,Hartnoll:2020fhc,Cai:2020wrp,VandeMoortel:2021gsp,Dias:2021afz,Cai:2021obq}, the matter fields generically destroy the inner horizon and the black hole develops matter hair.  Furthermore, the black hole singularity is  spacelike. 

As one falls into the black hole, a wealth of interesting phenomena with a remarkable dual interpretation occur (see again \cite{Hartnoll:2020rwq,Hartnoll:2020fhc,Cai:2020wrp,VandeMoortel:2021gsp,Dias:2021afz,Cai:2021obq}).  The purpose of this note is to describe the final, close-to-the-spacelike-singularity, stage of the dynamical evolution under generic initial condition (e.g., no assumed isometry or simplification leading to ``accidental'' behaviors).

The study of the asymptotic\footnote{In this paper, ``asymptotic'' always refers to the limit of going to the spacelike singularity,  which can be pushed to infinite time in a time coordinate adapted to the evolution. This limit is also called ``BKL limit'' and the spacelike singularity is often called ``cosmological singularity''. Similarly, ``generic'' means ``valid for an open set of intial data'' and thus stable under small deformations.} form of the generic solution of the Einstein equations near a spacelike singularity was pioneered in the seminal work by Belinski, Khalatnikov and Lifshitz (BKL) \cite{Belinsky:1970ew,Belinsky:1982pk}. This analysis was extended to general spacetime dimensions in \cite{Demaret:1985jnc,Demaret:1986ys}.  Massless $p$-forms (including $0$-forms) were dealt with in \cite{Belinski:1973zz,Belinsky:1981vdw,Damour:2000wm,Damour:2000hv,Damour:2002et}.  As shown in those works, the evolution of the dynamical fields at each spatial point can be described in terms of successive Kasner epochs, during which the fields follow a Kasner regime  (generalized to include the relevant scalar fields) characterized by definite Kasner exponents. A Kasner regime is stable and lasts all the way to the singularity if the  Kasner exponents fulfill some inequalities defining the ``Kasner stability region'', the specific form of which depends on the theory.  If the Kasner exponents are not in the stability region, the Kasner regime is replaced by another one with new Kasner exponents  before one reaches the singularity.  This replacement of the Kasner regime takes place through a rapid transition, called ``collision''.  The definite rules that give the new Kasner exponents in terms of the old ones can be computed in the asymptotic limit (i.e., sufficiently close to the singularity). 

Two competing behaviors can then occur as one goes to the singularity.   (I) Either there is no stability region and there is consequently an endless number of Kasner regimes, each characterized  by its own  Kasner exponents. This is the celebrated chaotic BKL oscillatory behavior, which holds in the original case of pure gravity in four spacetime dimensions, where it  is also named ``mixmaster'' following Misner who discovered it independently in the context of homogeneous ``Bianchi IX'' cosmological models \cite{Misner:1969hg}. (II)  Or there is a stability region and the system ends up in a Kasner regime with final Kasner exponents in that stability region. Although not the endless BKL oscillatory behavior, the BKL techniques of \cite{Belinsky:1970ew,Belinsky:1982pk} are perfectly adapted for handling this second case, which is in fact much simpler and for which analytical results can be rigorously established \cite{Andersson:2000cv,Damour:2002tc,Fournodavlos:2020tti,Fournodavlos:2020jvg,Ringstrom:2021ssc}.

For generic initial conditions, one can read off which asymptotic behavior will prevail  directly from the menu of fields and the Lagrangian (or equivalently and more easily, the Hamiltonian). 
Of course, even when chaos is generic (case I),  there always exists a set of measure zero of initial conditions for which the final asymptotic solution is of Kasner-for-ever type.  And conversely, if the system is not generically chaotic (case II), there might be a subset of solutions that exhibit the endless BKL oscillatory behavior.

An expedient and efficient way to determine whether it is case I or case II that occurs in a given theory is given by the billiard approach, which is asymptotically valid for a very general class of Lagrangians \cite{Damour:2000wm,Damour:2002et}.  The billiard approach is particularly interesting in revealing hidden symmetry structures \cite{Damour:2000hv}, but even when there is none, it is extremely powerful. 

We will assume that the reader is familiar with the billiard description of the dynamics near the singularity.  We refer to  \cite{Damour:2002et} for a detailed review, as well as to \cite{Henneaux:2007ej}
for complementary information and to \cite{Belinski:2017fas} for a review that covers also the original BKL approach. We shall only recall here the salient ideas.   

In the billiard description, the evolution of the dynamical fields at each spatial point is mapped on the motion of a ball in a portion of hyperbolic space.  The dimension of hyperbolic space and the walls bounding the billiard table are completely determined by the action. The Kasner regime corresponds to a geodesic motion in hyperbolic space.  This geodesic motion  is interrupted by bounces against the billiard walls, leading to new Kasner regimes.  In such a bounce, the transformation rules of Kasner exponents are just given by the standard specular reflection rules against the corresponding wall \cite{Damour:2000wm}.
The motion is chaotic (never-ending transitions) if the volume of the billiard table is finite.  It is non-chaotic, ending on a final Kasner regime if the volume of the billiard table is infinite, because there are in that case geodesic motions (with tangent vectors spanning an open set of directions) that never hit the walls and go unperturbed to infinity. 

The papers \cite{Belinski:1973zz,Belinsky:1981vdw,Damour:2000wm,Damour:2000hv,Damour:2002et} did not consider explicitly the above Lagrangians.
The purpose of this note is to show that the billiard description covers also (\ref{eq:LScalar})  and (\ref{eq:LVector}), i.e., that the mass terms (leading in the vector case to the absence of gauge invariance and a new, longitudinal degree of freedom) and the coupling terms can be incorporated into the picture. 

We then show that the relevant billiard has infinite volume for both the Lagrangians (\ref{eq:LScalar})  and (\ref{eq:LVector}), leading asymptotically to a final Kasner regime for both a massive charged scalar field and a massive charged vector field.  

One can understand the charged scalar field situation as follows.  The Einstein-Maxwell-neutral-scalar system has been much studied previously, allowing couplings between the vectors and the scalars of the exponential form $e^{ \lambda \phi} F_{\mu \nu} F^{\mu \nu}$ where $\lambda$ is the ``dilaton coupling''.  It was found that if the dilaton couplings belong to a well-defined ``subcritical region'' (which includes the origin) \cite{Damour:2002tc}, the evolution is non-chaotic and settles in a final Kasner regime, while if the dilaton couplings do not belong to that region, the evolution is chaotic and undergoes an infinite number of BKL oscillations.  In our case, $\lambda$ vanishes and is therefore in the subcritical region. As verified here, the minimal coupling terms of the charged scalar fields and the no-minimal coupling term $\alpha \phi^\dagger \phi  F_{\mu \nu} F^{\mu \nu}$ do not change that conclusion.

One can understand the vector field situation by observing that the massive vector field involves a longitudinal mode equivalent to a scalar field, with dilaton coupling again equal to zero and hence in the subcritical region. 
We stress that the presence of the longitudinal mode is crucial for reaching the conclusion of absence of chaos.  Without exciting it, one would find an infinite number of BKL oscillations typical of the Einstein-Maxwell system,  much in the same way as if one were to switch off the scalar field in the Einstein-scalar system. The longitudinal mode  is key, while the minimal couplings do not alter the conclusions.  [In that context, let us point out that the coupled Einstein-Yang-Mills system was explicitly treated in \cite{Belinski:2017fas}, following \cite{Belinski:1973zz,Belinsky:1981vdw}, and leads to conclusions on the asymptotic behavior identical to those holding for a collection of free massless vector fields.] 

\section{Charged black holes with scalar hair}

\subsection{Hamiltonian formulation}
We start with the case of a charged scalar field and consider first the absence of non-minimal coupling ($\alpha = 0$).  The (rescaled by $g^{\frac12}$) Hamiltonian constraint is in that case the sum of gravitational, Maxwell and scalar contributions,
\be
\mathcal{H} = \mathcal{H}^G + \mathcal{H}^{em} + \mathcal{H}^\phi \approx 0
\ee
with 
\begin{eqnarray}
 \mathcal{H}^G &=& G_{ijmn} \pi^{ij} \pi^{mn} - R g + 2 \Lambda g\\
  \mathcal{H}^{em} &=& \frac{1}{2} \pi^i \pi^j g_{ij} + \frac14 F_{ij} F_{mn} g^{im} g^{jn} g\\
   \mathcal{H}^\phi &=& \pi_\phi^\dagger  \pi_\phi + D_k \phi^\dagger D_m \phi g^{mn} g + m^2 \phi^\dagger \phi g
\end{eqnarray} 
where $g$ is the determinant of the spatial matric, $g^{ij}$ is its inverse, $G_{ijmn} = \frac12(g_{im} g_{jn} + g_{in} g_{jm}) - \frac{1}{d-1} g_{ij} g_{mn}$ with $d$ the number of spatial dimensions and $R$ is now the spatial curvature scalar.  The respective conjugate momenta are $\pi^{ij}$ (for the spatial metric $g_{ij}$), $\pi^i$ (for the spatial component of the electromagnetic vector potential $A_i$; $\pi^i$ is the electric field) and $\pi_\phi$, $\pi_\phi^\dagger$ (for the complex scalar field). The rescaled Hamiltonian constraint has density weight 2.  The contributions $\mathcal{H}^{em}$ and $\mathcal{H}^{\phi}$ are clearly definite positive.

There are other constraints in the theory, namely the momentum constraint and the Gauss constraint, but these are restrictions on initial data preserved by the evolution and so need only be imposed on the initial data.  The Hamiltonian constraint restricts of course also the initial data and do not differ from the other constraints in that respect, but it plays the additional role of generating the dynamics in the pseudo-Gaussian-temporal gauge defined by zero shift ($N^i =0$), rescaled lapse equal to unity ($\frac{N}{\sqrt{g}} = 1$) and $A_0 = 0$, in which $H = \int d^d x     \mathcal{H}$.  We take the pseudo-Gaussian coordinates to be adapted to the singularity, i.e., such that it occurs simultaneously everywhere in space. The time in that coordinate system is denoted by $\tau$, while the proper time is denoted by $t$ ($dt = \sqrt{g} d \tau$). With $\frac{N}{\sqrt{g}} = 1$, the singularity ($g \rightarrow 0$) occurs at $\tau \rightarrow + \infty$ \cite{Damour:2002et}. 

\subsection{Generalized Kasner metrics}
To get some insight into the system, let us first neglect the curvature, the electromagnetic contribution and keep only the kinetic energy of the scalar field, assuming that the metric is diagonal in some time-independent frame $\{l^i_a(x)\}$.  The Hamiltonian constraint governing the dynamics in that case reduces at each spatial point to \cite{Damour:2002et}
\be
\mathcal{H}_0 = \frac14 \left( \sum (\pi_a)^2 - \frac{1}{d-1}(\sum \pi_a)^2\right) + \pi_\phi^\dagger \pi_\phi \approx 0
\ee
where $\pi_a$ are the momenta conjugate to the logarithmic scale factors $\beta^a$ defined though $g_{aa} = e^{-2 \beta^a}$ (in the frame $\{l^i_a(x)\}$).  The equations of motion can easily be integrated and imply that $\pi_a$ and $\pi_\phi$ are constant, so that 
\be
\beta^a = v^a \tau + const, \qquad \phi = v^\phi \tau + const
\ee
in the gauge $\frac{N}{\sqrt{g}} = 1$, where $v^a$ and $v^\phi$ are integration constants that depend possibly on the spatial coordinates.

One then finds that $g = e^{-2 (\sum_a v^a) \tau}$. This implies $ \tau \sim - \ln \vert t \vert $ where we have fixed the origin of $t$ so that the (future) singularity is at $t=0$ (and approached from negative values of $t$).  One also gets  $\sqrt{g} \sim \vert t\vert $.

In terms of the proper time, the solution takes the generalized Kasner form\footnote{The metric is of a generalization of Kasner in that the Kasner parameters are inhomogeneous and scalar fields are included.}
\be
ds^2 = -dt^2 + \sum_{a=1}^d l_i^a l^a_j \vert t \vert^{2 p_a}dx^i dx^j, \qquad  \phi = -p_\phi \ln \vert t \vert + C_\phi
\ee
where  $C_\phi$ is an integration constant (depending on the spatial coordinates) and where the ``Kasner exponents'' $p_a, p_\phi$ (proportional to the velocities $v^a$, $v^\phi$ of $\beta^a$ and $\phi$)  fulfill
\be
\sum_{a = 1}^d p_a = 1, \qquad \sum_{a=1}^{d} p_a^2 + 4\vert p_\phi \vert ^2= 1, \label{eq:KasnerR}
\ee
the second condition following from the Hamiltonian constraint.  The solution has independent Kasner exponents at each spatial point.  One easily verifies that there is a curvature singularity at $t=0$ where $g \rightarrow 0$.  

Since the electromagnetic variables $(A_i, \pi^i)$ do not appear in $\mathcal{H}_0$, they obey $\dot{A}_i = 0$ and $\dot{\pi}^i = 0$ in the generalized Kasner solution and are frozen to some time-independent values. These integration constants can depend on the spatial coordinates.
The momentum and Gauss constraints imply conditions on the Kasner exponents, the $l^a_i$'s and the electromagnetic variables that are preserved by the evolution \cite{Belinsky:1970ew,Belinsky:1982pk,Damour:2002et,Belinski:2017fas}.

\subsection{Billiard dynamics}

We are now ready to analyse the effect  on the motion of the other terms in the Hamiltonian. 
As shown in \cite{Belinsky:1970ew,Belinsky:1982pk}, the asymptotic evolution involves a ``rotation of the Kasner axes'' $\{l^i_a(x)\}$ (see also  \cite{Belinski:2017fas}).  Furthermore, when $p$-forms are included, the metric generically does not remain diagonal. For that reason, it is convenient to analyse the dynamics in a different frame, the ``Iwasawa frame'' \cite{Damour:2002et} which has a group-theoretical meaning in terms of the homogeneous space $SL(d)/SO(d)$ and in which the description is simpler (the two frames coincide for diagonal metrics).  We shall thus parametrize a general  metric (with non-vanishing off-diagonal components) in terms of the Iwasawa variables $(\beta^a, \mathcal{N}^a_i)$ where $\beta^a $ are now the logarithmic scale factors in the Iwasawa frame and where the variables $\mathcal{N}^a_i$ parametrize the off-diagonal components. We also express the vector potential $A_i$ in the Iwasawa frame. The change of variables $(g_{ij}, A_i) \rightarrow (\beta^a, \mathcal{N}^a_i, A_a)$ is extended to the conjugate momenta so that it is a canonical transformation.

In Iwasawa frames, the Hamiltonian becomes at leading order in the limit of going to the singularity ($ g \rightarrow 0$) (see \cite{Damour:2002et} for details and justifications),
\be
\mathcal{H} = \mathcal{H}_0 + \mathcal{V}
\ee
where $\mathcal{H}_0 $ is the kinetic term for the momenta $\pi_a$ conjugate to the (logarithmic) scale factors  $\beta^a$ of the spatial metric in the Iwasawa frame and the scalar fields,
\be
\mathcal{H}_0 = \frac14 \left( \sum (\pi_a)^2 - \frac{1}{d-1}(\sum \pi_a)^2\right) + \pi_\phi^\dagger \pi_\phi
\ee
and where $ \mathcal{V}$ is a sum of infinite potential walls\footnote{These walls are actually of exponential type.  The exponentials  can be approximated by infinite step functions in the BKL limit \cite{Damour:2002et}.},   
\be
 \mathcal{V} = \sum_A \Theta(- 2 w_A (\beta^a)) 
\ee
with $ \Theta(x) = 0 $ for $x<0$ and $ \Theta(x) = \infty $ if $x>0$.

The wall forms $w_A(\beta)$ are linear forms in the $\beta^a$ originating from (i) the off-diagonal terms of the metric in $G_{ijmn} \pi^{ij} \pi^{mn}$ (``symmetry'', ``centrifugal'', or ``permutation'' walls); (ii) the spatial curvature $-R g$ (``curvature'' or ``gravitational'' walls); (iii) the electric energy density$ \frac{1}{2} \pi^i \pi^j g_{ij}$ (``electric'' walls); and (iii) the magnetic energy density $\frac14 F_{ij} F_{mn} g^{im} g^{jn} g$ (``magnetic walls'').  The cosmological constant, the potential energy density of the scalar fields and the mass term also brings walls, but these are irrelevant in the limit (see below).  The potential walls force the system to be in the region $w_A(\beta) \geq 0$.

The explicit form of the different types of walls is given in \cite{Damour:2002et,Belinski:2017fas}. These read in the case of the Lagrangian (\ref{eq:LTotal})-(\ref{eq:LScalar}) considered in this section:
\begin{itemize}
\item Symmetry walls, denoted $w_{(ab)} (\beta)$:
\be w_{(ab)}(\beta)=\beta^b - \beta^a, \qquad b > a \ee
As explained in \cite{Damour:2002et}, the collisions against the symmetry walls reorders the Kasner exponents and force the inequalities $\beta^1 \leq \beta^2 \cdots \leq \beta^d$. These walls are sometimes called permutation walls for that reason.
\item Curvature walls, denoted $\alpha_{abc}(\beta)$, with $a \not= b$, $a \not=c$ and $b \not=c$:
\be \alpha_{abc} (\beta) = 2 \beta^a + \sum_{e \not=a,b,c} \beta^e \ee
\item Electric walls $e_{a}(\beta)$:
\be e_a(\beta) = \beta^a \ee
\item Magnetic walls $m_a(\beta)$ ($a \not= b$):
\be m_{ab} (\beta) = \sum_{e\not= a, e \not=b} \beta^e \ee
\end{itemize}
The curvature term $-R g$ brings other walls $\mu_a(\beta)= \sum_{e \not= a} \beta^e$ but as shown in \cite{Damour:2002et,Belinski:2017fas}, these are irrelevant in the limit to the singularity because  the inequalities $\alpha_{abc} (\beta) \geq 0$ automatically imply $\mu_a (\beta) \geq 0$. 

It is clear that we need to keep only the ``dominant potential walls''  $w_{A'} (\beta)$ in the Hamiltonian, defined to be such that $w_{A'} (\beta) \geq 0$ imply all the inequalities $w_A(\beta) \geq 0$.  These are here:
\begin{itemize}
\item Symmetry walls:
\be \beta^2 - \beta^1, \quad \beta^3 - \beta^2, \cdots \quad \beta^d - \beta^{d-1} \ee
\item Electric wall:
\be e_1(\beta) = \beta^1 \ee
\end{itemize}
since $\beta^1 \geq 0$ and $\beta^{a+1} \geq \beta^a$ imply both $\beta^{a+n} \geq \beta^a$ and $\beta^a \geq 0$ for all $a$'s so that all symmetry, curvature, electric and magnetic wall forms are non negative. Note that in four spacetime dimensions ($d=3$), the magnetic wall $m_{23} (\beta)$ coincides with the electric wall $e_1(\beta)$ but in higher dimensions, the magnetic walls are subdominant.  The same is true for the curvature wall $\alpha_{123} (\beta)$ which reads $2 \beta^1$ in four spacetime dimensions and coincides thus with $e_1(\beta)$ up to the multiplicative factor $2$ but again, in higher dimensions, curvature walls are subdominant.

The reason that the cosmological constant can be neglected in the limit to the singularity was explained in section 6.4 of  \cite{Damour:2002et} and will not be repeated here.  That section explained also why a neutral  scalar term $ \partial_k \phi \partial_m \phi g^{mn} g$  could be neglected in the limit because $g^{mn} g$ brings the subdominant walls $\mu_a (\beta) = \sum_{e \not= a} \beta^e$,  so that $\partial_k \phi \partial_m \phi g^{mn} g$ goes to $\sum_a C_a\Theta(- 2 \mu_a(\beta))$.  The (non negative) prefactors $C_a$ of the $\Theta$-function depend on the scalar field  and can be absorbed in $\Theta(- 2 \mu_a(\beta))$, which is either $0$ or $\infty$ (the scalar field does not overcome this behavior \cite{Damour:2002et}).  So, the term $\partial_k \phi \partial_m \phi g^{mn} g$ becomes $\sum_a \Theta(- 2 \mu_a(\beta))$ and since the walls $\mu_a(\beta)$ are subdominant, this term can be dropped with respect to the dominant wall potential terms. 
The same argument  implies  that $D_k \phi^\dagger D_m \phi g^{mn} g$ (with covariant derivatives) is equally negligeable in the limit because $D_m \phi$ differs from $\partial_m \phi$ by $- i q A_m \phi$, but $A_m$ freezes to a constant in the limit \cite{Damour:2002et} so that again, all prefactors of the $\Theta$-functions can be absorbed, leading to the same subdominant-wall expression $\sum_a \Theta(- 2 \mu_a(\beta))$that can be neglected.  That the mass term is also crushed asymptotically, as the cosmological term,  is a consequence that it is mutiplied by $g$ which goes to zero.  The coefficient $\phi^2$  blows up but much slowlier and thus the product goes to zero.

\subsection{Final Kasner regime}

Since the mass term and the minimal coupling terms of the scalar field can be neglected as one goes to the singularity, the discussion of the asymptotic behaviour of the system is the same as the one for the coupled Einstein-Maxwell system with two neutral scalar fields. The ``dilaton coupling'' $\lambda$ vanishes  for the Lagrangian (\ref{eq:LTotal})-(\ref{eq:LScalar}) since the kinetic term for the vector field reads $(-1/4) F_{\mu \nu} F^{\mu \nu} = (-1/4) e^{0. \phi} F_{\mu \nu} F^{\mu \nu}$.

This system has already been much studied.  Because the dilaton coupling $\lambda$ is in the subcritical region \cite{Damour:2002tc}, the system generically undergoes at most a finite number of oscillations before settling down to a final Kasner regime\footnote{Note that Kaluza-Klein reduction of the pure Einstein theory crucially induces non-vanishing dilaton couplings, which can change the picture.}.

The reasoning goes as follows.  The billiard ball representing the independent logarithmic scale factors  and the scalar fields at each point moves in a region of hyperbolic space of dimension $d + 1$ (there are $d$ logarithmic scale factors and $2$ real scalar fields, and one relation following from the Hamiltonian constraint). The number of independent relevant walls is $d$ ($d-1$ symmetry walls and $1$ electric wall), which is insufficient to bound a finite-volume region in hyperbolic space. Hence, there are ``escape directions'' to infinity forming an open set, in which the final Kasner regime can settle\footnote{This is the description obtained by projecting the motion to the upper sheet of the unit hyperboloid  in the space of the logarithmic scale factors $\beta^a$  and the scalar fields $\phi$ factors.  This space has a Minkowskian structure.  The unprojected motion in that space is a broken lightlike straight line, with collisions against the wall hyperplanes, which are timelike.  If one focusses only on the gravitational scale factors $\beta^a$, which is permitted because the velocities of the scalar fields do not change under collisions (the wall forms do not involve the scalars), one finds that because of the scalar field contribution to $\mathcal{H}_0$, these move along  broken timelike straight lines.  There are clearly open cones of timelike directions that avoid the wall hyperplanes.  For such an interrupted motion, the solution is of Kasner-for-ever type.  More information in  \cite{Damour:2002et}.}.  

This final Kasner regime( $\tau \rightarrow \infty$)  is such that all velocities $v^a$ are  positive so that $v^a \tau$ and hence $\beta^a$ is positive (for large $\tau$).  Furthermore, the $v^a$'s are ordered ($v^1 \leq v^2 \leq \cdots \leq v^d$) so that the symmetry wall forms are also positive.  That these conditions can be fulfilled follows from the fact  that the Kasner relations (\ref{eq:KasnerR}) allow strictly positive Kasner exponents $p_a$ (proportional to the velocities) when the scalar fields are present.

Another way to see the same thing goes as follows. On the generalized Kasner solution (with scalar fields), the individual terms in $\mathcal{H}_0$ are of order $1$. The difference between the exact Hamiltonian $\mathcal{H}$ and the unperturbed Hamiltonian $\mathcal{H}_0$  involves only terms of the form $t^S$ (possibly multiplied by coefficients involving $(\ln t)^k$), where the $S$'s are sums of Kasner exponents with positive coefficients.  When the Kasner exponents $p_a$ are all strictly positive, the $S$'s are also strictly positive so that $(\ln t)^k t^S \rightarrow 0$ as $t \rightarrow 0$.  The correction terms vanish and the generalized Kasner solution is thus asymptotically exact.  

From that perspective, the cosmological constant term goes as $t^2$ while the scalar potential term $D_k \phi^\dagger D_m \phi g^{mn} g$ goes as $(\ln \vert t \vert )^2 t^{P}$ with the minimal coupling terms yielding   $\ln \vert t \vert  t^{P}$ and $t^{P}$, while the mass term goes as $(\ln \vert t \vert )^2 t^2$.  Here, the exponents $P$ are also sums of Kasner exponents with positive coefficients.  All these terms are thus indeed negligeable in the limit.  Note incidentally that the $P$'s are actually greater than the $S$'s associated with the dominant walls. [A detailed review of the BKL perturbation expansion near the singularity is given in Section 1.6 of \cite{Belinski:2017fas}.]

Before settling into the final Kasner regime, the system  undergoes a finite number of collisions.  The transition rule from the old to the new Kasner exponents in a collision takes asymptotically a simple form, when the approximation of the exponential potentials by infinite wall potentials is valid.  However, since the last collision occurs a finite proper time away from the singularity, this simple rule might not be an accurate approximation if one has not entered yet the asymptotic regime.  This will depend on the initial conditions.  Understanding the pre-asymptotic regime is a rather intricate and interesting question for which we refer to \cite{Hartnoll:2020rwq,Hartnoll:2020fhc,Cai:2020wrp,VandeMoortel:2021gsp,Dias:2021afz}.

As we have seen, the infinite step functions defining the walls come with prefactors that depend on the fields.  When these prefactors are non-zero, they can be absorbed in the infinite step functions. However, it may happen that for some particular field configurations (forming a set of measure zero), one (or more) prefactor vanishes. This could occur when a particular symmetry is imposed.  In that case, the corresponding wall is absent and the billiard table is bigger than the one relevant for generic initial conditions.   If the volume of the billiard table is already infinite when all walls are included, as here, the removal of a wall will not change the qualitative behavior of the solutions, but it might have a more dramatic impact when the generic billiard table has finite volume.
 
It should also be noted that while the asymptotic Kasner behavior  is the generic situation, there exits a subset of measure zero of solutions that do exhibit a never-ending chaotic BKL oscillatory:  simply set the scalar fields equal to zero. Chaos is driven when $\phi=0$ by the electric walls, which are the dominant walls.  We thus see that the scalar fields play a crucial role in the conclusions.   Examining a subset of measure zero of solutions cannot be used to safely draw conclusions valid in the generic case.

\subsection{Non-minimal couplings}

We now include the non-minimal term $- \alpha \phi^\dagger \phi  F_{\mu \nu} F^{\mu \nu}$.  This term introduces two modifications in the Hamiltonian:
\begin{enumerate}
\item First the conjugate momentum $\pi^i$ to $A_i$ receives a contribution from  $-2 \alpha \phi^\dagger \phi  F_{0 n} F^{0 n}$, leading to the following expression for the electric energy density (including the Maxwell contribution),
\be
 \frac{1}{2 (1 + 4 \alpha \phi^\dagger \phi)} \pi^i \pi^j g_{ij}
 \ee
\item Second, the total magnetic energy is modified as
\be
 \frac{(1 + 4 \alpha \phi^\dagger \phi)}{4} F_{ij} F_{mn} g^{im} g^{jn} g 
 \ee
\end{enumerate}
The coefficient $\alpha$ is assumed to be positive so that the above energy density terms are positive \cite{Dias:2021afz}.

The non-minimal terms therefore only modify the prefactors of the electric and magnetic wall potentials (without changing their sign or making them equal to zero) and can be absorbed in the BKL limit. They have therefore been already taken into account and do not bring new features asymptotically.  After a finite number of collisions, 
the solution will again settle in a final Kasner regime characterized by positive, ordered, Kasner exponents $p_a$. The terms introduced by the non-minimal couplings are as above of the form $t^S (\ln t)^k$ with $S>0$ and hence go to zero in the limit. 

We stress that this is radically different from exponential scalar (dilaton) couplings $\sim e^{\lambda \phi}F_{\mu \nu} F^{\mu \nu}$ which do have a on trivial impact since they modify the electric and magnetic wall forms, and hence, the billiard region \cite{Damour:2002et}.

\section{Charged black holes with vector hair}

\subsection{Hamiltonian formulation}
The Hamiltonian constraint reads in the charged vector case
\be
\mathcal{H} = \mathcal{H}^G + \mathcal{H}^{em} + \mathcal{H}^{(\rho)} \approx 0
\ee
with 
\begin{eqnarray}
 \mathcal{H}^G &=& G_{ijmn} \pi^{ij} \pi^{mn} - R g + 2 \Lambda g\\
  \mathcal{H}^{em} &=& \frac{1}{2} \pi^i \pi^j g_{ij} + \frac14 F_{ij} F_{mn} g^{im} g^{jn} g\\
   \mathcal{H}^{(\rho)} &=& P^{i\dagger}  P^j g_{ij} +\frac12 \rho_{ij}^\dagger \rho_{mn} g^{im} g^{jn} g + m^2 \rho_i^\dagger \rho_j g^{ij}  g + \frac{1}{m ^2} D_i P^{i \dagger} D_j P^j
\end{eqnarray} 
where $P^{i \dagger}$ and $P^i$ are the momenta conjugate to $\rho_i$ and $\rho_i^\dagger$, respectively.   The variables $\rho_0$ and $\rho^\dagger_0$ have been eliminated through their own equations of motion, which is possible when $m^2 \not=0$ (no gauge invariance).   Since $m^2 >0$ the various terms in $\mathcal{H}^{(\rho)}$ are positive.  The momentum and Gauss constraints receive also contributions from the charged vector field $\rho_i$ and its conjugate, which transform under the corresponding gauge symmetries, but these will not concern us here since again, they can be taken into account by imposing them on the initial data.  In the pseudo-Gaussian coordinate system (with $A_0 =0$), the equations of motion are generated by $H= \int d^d x \mathcal{H}$.  

In the BKL limit, the dynamical evolution is now controlled by 
\be
\mathcal{H} = \mathcal{H}_0 + \mathcal{V}
\ee
where $\mathcal{H}_0 $ is the kinetic term for the momenta $\pi_a$ conjugate to the (logarithmic) scale factors  $\beta^a$ of the spatial metric in the Iwasawa frame  {\it and}  the momenta conjugate to the longitudinal modes
\be
\mathcal{H}_0 = \frac14 \left( \sum (\pi_a)^2 - \frac{1}{d-1}(\sum \pi_a)^2\right) + \frac{1}{m ^2} D_a P^{a \dagger} D_a P^a
\ee
and where $ \mathcal{V}$ is a sum of infinite potential walls
\be
 \mathcal{V} = \sum_A \Theta(- 2 w_A (\beta^a)) 
\ee
with:
\begin{itemize}
\item Symmetry walls:
\be w_{(ab)}(\beta)=\beta^b - \beta^a, \qquad b > a \ee
\item Curvature walls:
\be \alpha_{abc} (\beta) = 2 \beta^a + \sum_{e \not=a,b,c} \beta^e \qquad (a \not= b, a \not=c ,b \not=c)\ee
\item Electric walls $e_{a}(\beta)$:
\be e_a(\beta) = \beta^a \ee
\item Magnetic walls $m_a(\beta)$ ($a \not= b$):
\be m_{ab} (\beta) = \sum_{e\not= a, e \not=b} \beta^e \ee
\item Electric walls of the charged vector field $E_{a}(\beta)$:
\be E_a(\beta) = \beta^a \ee
\item Magnetic walls of the charged vector field $M_{ab} (\beta)$ ($a \not= b$):
\be M_{ab} (\beta) = \sum_{e\not= a, e \not=b} \beta^e \ee
\end{itemize}

The term $\frac{1}{m ^2} D_a P^{a \dagger} D_a P^a$ is included in $\mathcal{H}_0$ because it is of the same order $O(1)$ as the other terms in $\mathcal{H}_0$  since it does not involve the metric. Not including it in $\mathcal{H}_0$ would lead to an immediate contradiction  because the equations of motion derived from $\mathcal{H}_0$ imply that the $P^a$'s are time independent and hence $\frac{1}{m ^2} D_a P^{a \dagger} D_a P^a$ is of order $O(1)$.  The same conclusion that $\frac{1}{m ^2} D_a P^{a \dagger} D_a P^a$ is of order $O(1)$ follows also from the complete Hamiltonian, because the only potential term involving the longitudinal modes is $m^2 \rho_i^\dagger \rho_j g^{ij}  g$ but this term can be neglected asymptotically since the metric factor $g^{ij}g$  becomes equivalent to a potential wall potential of the form $\Theta(- 2\mu_a(\beta))$, which is subdominant, and can be replaced by zero in the limit.  

The longitudinal modes therefore parallel exactly  the scalar fields.

\subsection{Final Kasner regime} 

Because the longitudinal modes can be assimilated to scalar fields, one can draw the same conclusions: the system will asymptotically settle to a final Kasner regime for generic initial conditions.  At the same time, the transverse terms of the massive charged vector field simply duplicate the walls already brought by the electromagnetic field and do not change the billiard shape. 

Perhaps the best way to emphasize the fact that the longitudinal mode is equivalent to a scalar field is to introduce a Stueckelberg scalar field $\phi$ through the replacement of $\rho_\mu$ by $\rho_\mu - \partial_\mu \phi$, which makes the theory gauge invariant under $\rho_\mu \rightarrow \rho_\mu + \partial_\mu \epsilon$, $\phi \rightarrow \phi + \epsilon$.  The kinetic term of the Stueckelberg scalar field is included in $\mathcal{H}_0 $ as for an ordinary scalar field.

In order to identify the generic behavior of the system, it is crucial to allow a non-vanishing  longitudinal component.  In the very interesting paper \cite{Cai:2021obq}, the behavior of the model  described by the action (\ref{eq:LVector}) was analyzed in the singularity  limit and it was concluded that the chaotic oscillatory behavior prevailed.   This conclusion is correct in the non-generic context considered there, which had vanishing longitudinal component. The $\rho$-field of the solutions of \cite{Cai:2021obq} was indeed assumed to have only one spatial component depending only on time (denoted $z$ close to the singularity in \cite{Cai:2021obq}).  The resulting vector field is spatially homogeneous and  has zero divergence $D_a P^a$. The system is then asymptotically described by the Einstein-Maxwell theory, with three vector fields. [The well-known fact that chaos is driven then by the electric walls of the electromagnetic field and of the vector field $\rho_\mu$ , which dominate the magnetic and the curvature walls, and the well-known reflection rule against these walls \cite{Damour:2002et,Belinski:2017fas}, were also observed in \cite{Cai:2021obq}.] Re-establing the longitudinal mode  destroys chaos, in the same way as  switching on a scalar field. 

It is actually satisfactory that this is so, since it means that introducing a mass either  ``\'a la Proca''  or through the Brout-Englert-Higgs mechanism with an explicit scalar field lead to identical conclusions.

\section{Conclusions}
\label{sec:Conclusions}

In this paper, we have investigated the behavior near a spacelike singularity of the coupled Einstein-Maxwell-(charged) scalar and Einstein-Maxwell-(charged and massive) vector systems.  We have shown that because of the scalar field in the first case, and of the  longitudinal mode of the massive vector field in the second case, the systems both settle to a final Kasner regime.  

Since this monotonous final regime is simpler than the oscillatory chaotic one, it is expected that rigorous analytical results can be established along the lines of \cite{Andersson:2000cv,Damour:2002tc,Fournodavlos:2020tti,Fournodavlos:2020jvg,Ringstrom:2021ssc}.  It would be of interest to do so.

Finally, we briefly comment on the non-minimal coupling term $iq \gamma \rho_\mu \rho_\nu^\dagger F^{\mu \nu}$ that can be added to the Lagrangian of the vector case.  This term does not  seem to lead to an energy density which is bounded from below.  To explore its role near the singularity, one must analyse configurations for which  $\rho_\mu \rho_\nu^\dagger F^{\mu \nu} \not=0$.
The BKL impact of this term appears to be intricate because it modifies the momentum conjugate to the electromagnetic vector potential. A preliminary study seems to indicate that its ultimate effect is to suppress the kinetic term $\frac{1}{m ^2} D_a P^{a \dagger} D_a P^a$.  If so, the non-minimal coupling term $iq \gamma \rho_\mu \rho_\nu^\dagger F^{\mu \nu}$ would lead to the oscillatory behavior.  More analysis is required to draw definite conclusions.

\section*{Acknowledgments}
This work was partially supported by FNRS-Belgium (conventions FRFC PDRT.1025.14 and IISN 4.4503.15), as well as by funds from the Solvay Family.

%%%*************************************

%\appendix
%dummy comment inserted by tex2lyx to ensure that this paragraph is not empty%dummy comment inserted by tex2lyx to ensure that this paragraph is not empty%dummy comment inserted by tex2lyx to ensure that this paragraph is not empty

%\section{Transformation laws of the subleading terms} \label{App1}
%AAA

%%***************************************************

\end{document}